\documentclass[12pt]{article}
\usepackage{graphicx}

\usepackage{url}
\usepackage{color, colortbl}
\definecolor{lightblue}{rgb}{0.93,0.95,1.0}
\definecolor{LightCyan}{rgb}{0.88,1,1}
\definecolor{lightgray}{rgb}{0.83, 0.83, 0.83}
\definecolor{pink}{rgb}{0.82, 0.52, 0.5}

\def\UCSD{Department of Physics, University of California San Diego, La Jolla, CA, USA;\\}

\def\Title#1{\begin{center} {\Large #1 } \end{center}}
\def\Author#1{\begin{center}{ \sc #1} \end{center}}
\def\Address#1{\begin{center}{ \it #1} \end{center}}

\newenvironment{Presented}{\begin{quotation} \begin{center} 
             PRESENTED AT\end{center}\bigskip 
      \begin{center}\begin{large}}{\end{large}\end{center} \end{quotation}}

\def\beq{\begin{equation}}
\def\eeq#1{\label{#1}\end{equation}}
\def\eeqn{\end{equation}}

\def\beqa{\begin{eqnarray}}
\def\eeqa#1{\label{#1}\end{eqnarray}}
\def\eeqan{\end{eqnarray}}

\let\bar=\overbar

\def\Dslash{\not{\hbox{\kern-4pt $D$}}}
\def\dslash{\not{\hbox{\kern-2pt $\del$}}}

\def\msb{{\bar{\ssstyle M \kern -1pt S}}}

\begin{document}
\begin{titlepage}
%\pubblock

\vfill
\Title{The Simons Observatory: Project Overview}
\vfill
\Author{Nicholas Galitzki on behalf of the Simons Observatory Collaboration}
\Address{\UCSD}
\vfill
\begin{abstract}
The Simons Observatory (SO) will make precision temperature and polarization measurements of the cosmic microwave background (CMB) over angular scales between 1 arcminute and tens of degrees using over 60,000 detectors and sampling frequencies between 27 and 270 GHz. SO will consist of a six-meter-aperture telescope coupled to over 30,000 detectors and an array of half-meter aperture refractive cameras, coupled to an additional 30,000+ detectors. The unique combination of large and small apertures in a single CMB observatory will allow us to sample a wide range of angular scales over a common survey area while providing an important stepping stone towards the realization of CMB-Stage IV. CMB-Stage IV is a proposed project that will combine and expand on existing facilities in Chile and Antarctica to reach the ~500,000 detectors required for CMB-Stage IV's science objectives. SO and CMB-Stage IV will measure fundamental cosmological parameters of our universe, constrain primordial fluctuations, find high redshift clusters via the Sunyaev-Zeldovich effect, constrain properties of neutrinos, and trace the density and velocity of the matter in the universe over cosmic time. The complex set of technical and science requirements for SO has led to innovative instrumentation solutions which we will discuss. For instance, the SO large aperture telescope will couple to a cryogenic receiver that is 2.4\,m in diameter and 2.4\,m long. We will give an overview of the drivers for and designs of the SO telescopes and cameras as well as the current status of the project. We will also discuss the current status of CMB-Stage IV and important next steps in the project's development.

\end{abstract}
\vfill
\begin{Presented}
Thirteenth Conference on the Intersections of Particle and Nuclear Physics\\
Palm Springs, CA,  May 29 - June 3, 2018
\end{Presented}
\vfill
\end{titlepage}
\def\thefootnote{\fnsymbol{footnote}}
\setcounter{footnote}{0}
%

%%%%%%%%%%%%%%%%%%%%%%%%%%%%%%%%%%%%%%%%%%%%%%%%%%%%%%%%%%%%%
\section{Overview}
\label{sec:intro}  % \label{} allows reference to this section

The cosmic microwave background (CMB) has emerged as one of the most powerful probes of the early universe. Measurements of temperature anisotropies on the level of ten parts per million have brought cosmology into a precision era, and have placed tight constraints on the fundamental properties of the Universe\cite{Planck2018}. Beyond temperature anisotropies, CMB polarization anisotropies not only enrich our understanding of our cosmological model, but could potentially provide clues to the very beginning of the universe via the detection (or non-detection) of primordial gravitational waves. To provide a complete picture of cosmology, measurements at multiple frequencies of both large and small angular scales are important.  

\begin{figure}[t]
    	\begin{center}
        \begin{tabular}{c}
        \includegraphics[width = 1.0\linewidth]{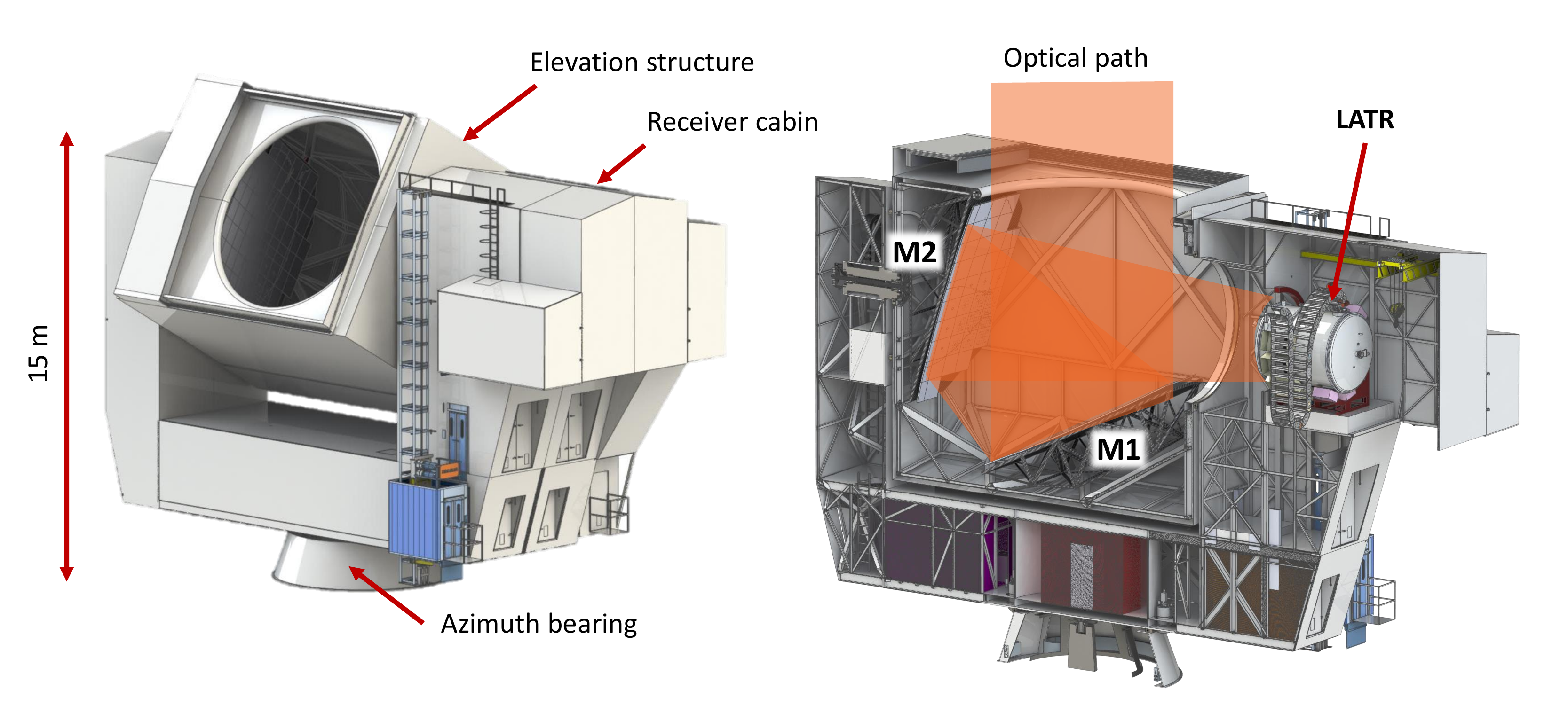}
        \end{tabular}
        \end{center}
    	\caption{A cross-section showing how the LATR couples to the LAT.  In the orientation shown on the right, light enters the telescope from the top.  It then reflects off the 6\,m primary (M1) and 6\,m secondary (M2) before being directed to the receiver. The receiver always operates in the horizontal orientation shown here; it is not directly mechanically coupled to the telescope elevation structure.  Therefore, a separate mechanism is used to rotate the receiver about its long axis as the telescope elevation structure moves M1 and M2 in rotation.
        %http://simonsobservatory.wdfiles.com/local--files/tech%3Aopticaldesign/Optics_tube_receiver_layouts_20170616.pdf?ukey=e7a71d7bbb0533625ec36e0a6f8dde100f94016b
        %http://simonsobservatory.wdfiles.com/local--files/tech%3Aopticaldesign/SO_telescope_summary_20170813.pdf?ukey=68ad5a0279f58a1f3d8dda7bc4480dc8665d9757
        %http://simonsobservatory.wikidot.com/lenssizenotes
        } 
        \label{fig:telescope}
    \end{figure}

\begin{figure}[t]
  \centerline{
    \includegraphics[height=3.0in]{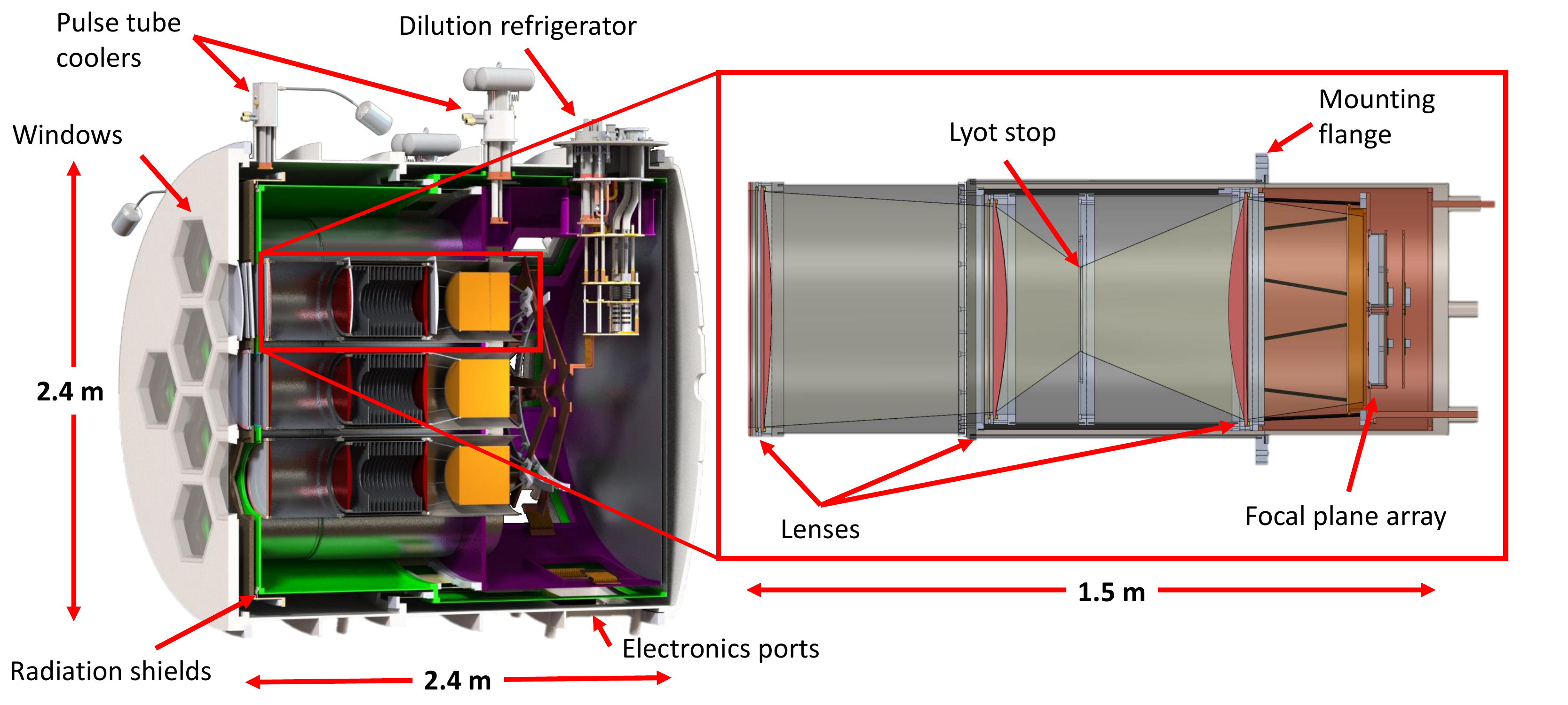}
  }
  \caption{Cross sectional view of the LATR with primary components labeled. A cross section of a single optics tube is detailed at right with the main optical components labeled.
    \label{fig:latr_1}}
\end{figure}

SO will field a 6\,m diameter crossed Dragone\cite{Dragone1978} large aperture telescope (LAT) coupled to the large aperture telescope receiver (LATR). The LAT is designed to have a large field of view (FOV)\cite{Niemack2016,Parshley2018}, 7.2$^\circ$ at 90 GHz (Fig. \ref{fig:telescope}).  
During the initial deployment, we plan to populate seven out of thirteen total optics tubes, each containing three lenses and three detector wafers, for a total of over 30,000 detectors. The optics tubes are modular and capable of being inserted from the back of the cryostat while it is mounted on the telescope. The refractive reimaging optics for each optics tube are composed of three silicon lenses with machined metamaterial antireflection (AR) surface layers\cite{Datta2013}. 
An overview of the receiver and optics tube design is shown in Fig. \ref{fig:latr_1}.
%~~~~~~~~~~~~~~~~~~~~~~~~~~~~
%~~~~~~~~~~~~~~~~~~~~~~~~~~~~

SO will also deploy an array of 42 cm aperture small aperture telescopes (SATs) with seven detector wafers in each coupled to over 30,000 detectors total. SO will observe with three frequency pairs in order to observe the CMB peak signal and constrain polarized foreground contamination from galactic synchotron and dust emission at lower and higher frequencies. The SO frequency bands are: 27/39 GHz, low frequency(LF); 90/150 GHz, mid-frequency(MF); and 220/270 GHz, ultra-high frequency(UHF). The LATR will initially be populated with four MF optics tubes, two UHF optics tubes, and one LF optics tube, while the SAT array will initially be composed of two MF SATs and one UHF SAT with a fourth LF SAT to follow.

More details of the SO instrumental configuration can be found in Galitzki et al. 2018\cite{Galitzki2018} as well as an upcoming instrument overview paper. More details of the SO science objectives can be found in `The Simons Observatory: Science goals and forecasts.'\cite{SOSci}

A description of the CMB-S4 science case can be found in the `CMB-S4 Science Book, first edition'\cite{s4sci} while a description of CMB-S4 technology development is described in the `CMB-S4 Technology Book, first edition.'\cite{s4tech}

%%%%%%%%%%%%%%%%%%%%%%%%%%%%%%%%%%%%%%%%%%%%%%%%%%%%%%%%%%%%%
\section{Sensor technology}
\label{sec:sensors}

%~~~~~~~~~~~~~~~~~~~~~~~~~~~~
%\input{Figures/UFMFigure}
%~~~~~~~~~~~~~~~~~~~~~~~~~~~~

SO uses AlMn transition-edge sensor (TES)\cite{Li2016} bolometers fabricated on 150\,mm diameter silicon wafers with two demonstrated technologies for radio frequency (RF) coupling: lenslet coupled sinuous antennas\cite{Suzuki2012} and horn coupled orthomode transducers (OMTs)\cite{Mcmahon2012, Choi2018, Simon2016, Henderson2016}.  The bolometer arrays with sinuous antennas will be fabricated at the University of California - Berkeley (UCB) while the ones with OMTs will be fabricated at the National Institute of Standards and Technology (NIST)\cite{Duff2016}. The TESs are read out with microwave SQUID multiplexing ($\mu$mux) \cite{Mates2011, Dober2017} which has been demonstrated in the MUSTANG2 camera\cite{Dicker2014}. Each detector pixel has four bolometers to sense two orthogonal polarizations in each of the two frequency bands. The TESs are tuned to have a $T_c\approx160$\,mK. 

\subsection{Microwave multiplexing readout electronics}

The $\mu$mux readout technology is capable of reading out thousands of detectors on a single pair of RF lines. The high multiplexing factor vastly simplifies the cabling for the detectors which must penetrate a vacuum shell and multiple radiation shields. Additional details of the $\mu$mux system can be found in Dober et al. 2017\cite{Dober2017} and Henderson et al. 2018\cite{Henderson2018}.

%~~~~~~~~~~~~~~~~~~~~~~~~~~~~
%========================================================%
%========================================================%
%~~~~~~~~~~~~~~~~~~~~~~~~~~~~

%%%%%%%%%%%%%%%%%%%%%%%%%%%%%%%%%%%%%%%%%%%%%%%%%%%%%%%%%%%%%

\section{Small aperture telescope}
\label{sec:sat}

%~~~~~~~~~~~~~~~~~~~~~~~~~~~~
   \begin{figure}[t]
    	\begin{center}
        \includegraphics[width = 1.0\linewidth]{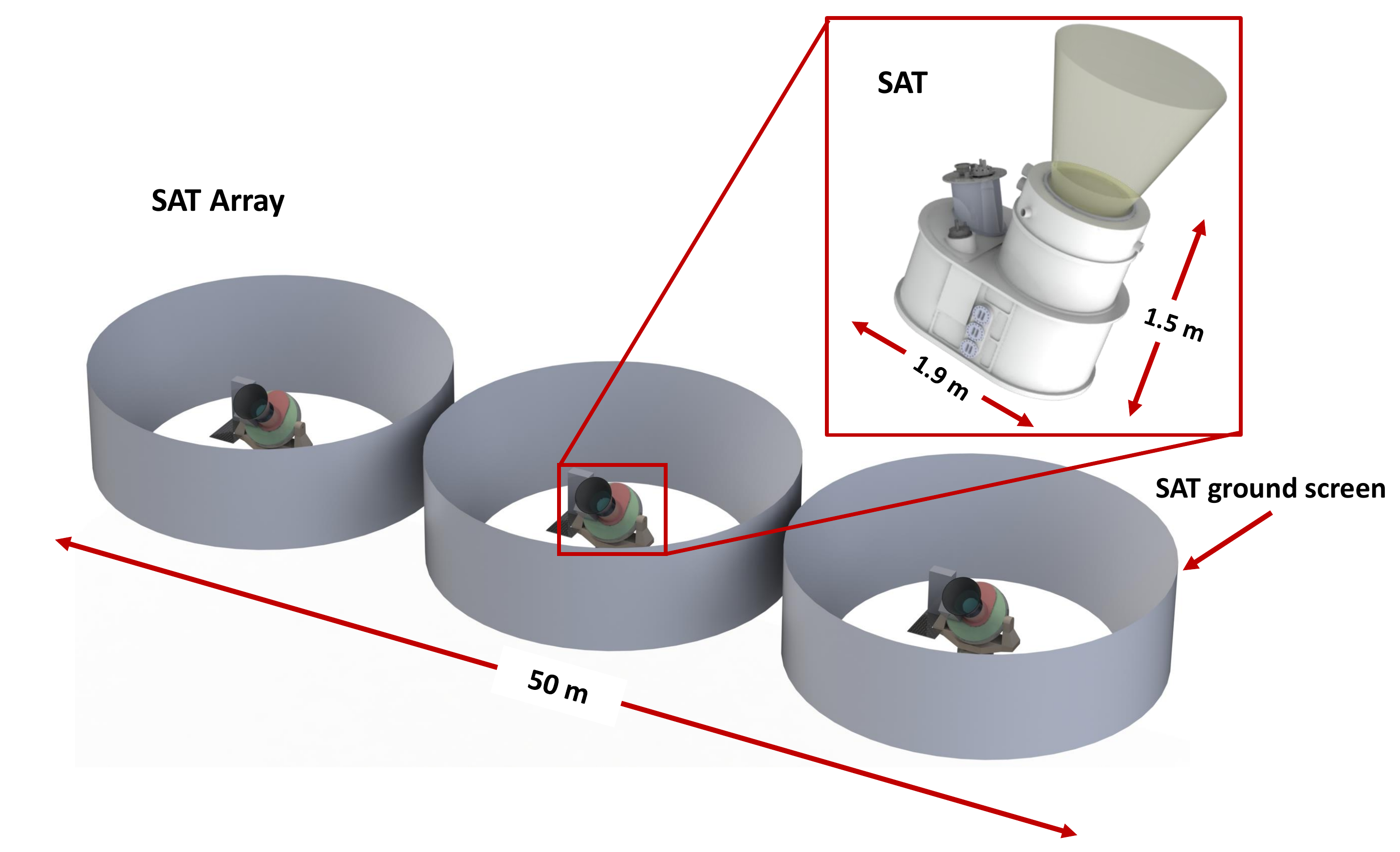}
        \end{center}
    	\caption[Caption for SAT Array]{A conceptual image of the SAT array consisting of three telescopes on pointing platforms with each surrounded by a ground shield. The inset shows a single SAT receiver with the rays from the 35$^\circ$ FOV beam shown as the shaded cone.
} 
        \label{fig:satarr}
    \end{figure}
%~~~~~~~~~~~~~~~~~~~~~~~~~~~~
The SO SAT is designed to observe CMB polarization signals on degree angular scales, where a faint peak in the parity-odd polarization signal, over four orders of magnitude smaller than CMB temperature anisotropies, is predicted to occur. The SAT science goals require a deep survey area with impeccable systematics control.
Optimizing the throughput of the SAT provided the simplest path to maximizing mapping speed which resulted in a short focal length system with a large FOV (35$^\circ$) that couples to seven 150\,mm detector wafers. The SAT design increases mapping speed by cooling the aperture stop and lenses to 1\,K. The SAT receiver will also house a continuously rotating cryogenic half-wave plate (CHWP) between the receiver window and first lens to modulate the polarized signal to allow for additional systematics control.

\subsection{SAT cryogenic receiver}

%~~~~~~~~~~~~~~~~~~~~~~~~~~~~
%========================================================%
\begin{figure}[t]
  \centerline{
    \includegraphics[height=3.0in]{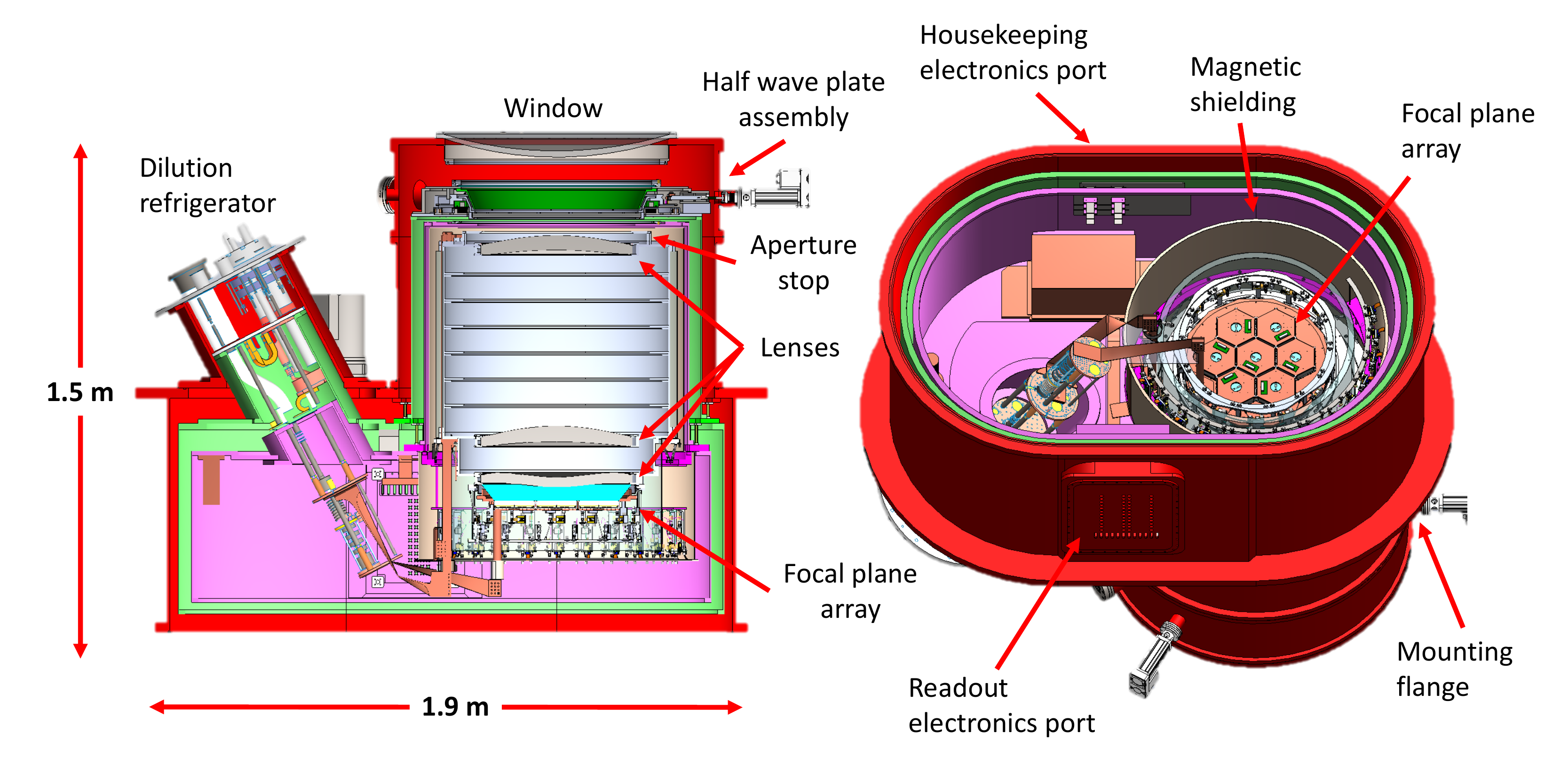}
  }
  \caption{Detailed views of the primary components of the SAT receiver. The image at left shows a cross-sectional view highlighting the optics tube and focal plane array placement as well as the angled dilution refrigerator. The view at right shows the back end of the receiver with the lid section cut away showing the primary components as they are situated in the 4\,K volume. 
    \label{fig:sat}}
\end{figure}
%========================================================%
%~~~~~~~~~~~~~~~~~~~~~~~~~~~~

 The SAT receiver is composed of a primary cylindrical volume surrounding the three silicon lens refractive optics and focal plane combined with a secondary volume to accommodate the cryogenic systems as shown in Fig. \ref{fig:sat}. The CHWP requires an additional space in front of the optics tube for mounting and operation. The readout systems for the detectors also require a dedicated electronics port in the side of the cryostat that presented a significant design challenge to ensure appropriate thermal coupling, cable routing, and component mounting for the readout chain operational requirements.

%%%%%%%%%%%%%%%%%%%%%%%%%%%%%%%%%%%%%%%%%%%%%%%%%%%%%%%%%%%%%
\section{Conclusion}

We have presented an overview of the principal components of SO as well as their current status and the design choices that led to our final instrument configuration. SO is on target to begin scientific observations of the CMB with the LAT and SAT starting in the year 2021. SO will be one of the most sensitive broadband, wide angular scale, CMB survey instruments to date. With its initial deployment of over 60,000 detectors, SO will provide an important step forward in CMB science and pave the way for future millimeter wave experiments such as CMB-S4\cite{s4sci, s4tech}.

%%%%%%%%%%%%%%%%%%%%%%%%%%%%%%%%%%%%%%%%%%%%%%%%%%%%
%\appendix    %>>>> this command starts appendixes

%%%%%%%%%%%%%%%%%%%%%%%%%%%%%%%%%%%%%%%%%%%%%%%%%%%%%%%%%%%%%
%\acknowledgments     %>>>> equivalent to \section*{ACKNOWLEDGMENTS}
This work was supported in part by a grant from the Simons Foundation (Award \#457687, B.K.)

\bibliography{report}   %>>>> bibliography data in report.bib
\bibliographystyle{spiebib}   %>>>> makes bibtex use spiebib.bst

\end{document}